\documentclass[aps,prl,twocolumn,showpacs,showkeys,superscriptaddress]{revtex4-1}
\usepackage{graphicx}% Include figure files

\newcommand{\met}{\mbox{${\not\!\!E_{\rm T}}$}}
\newcommand{\metvec}{\mbox{${\not\!\! \vec{E}_{\rm T}}$}}

\newcommand{\fb}{\ensuremath{\mathrm{fb}^{-1}}}
\newcommand{\KeV}{\ensuremath{\mathrm{Ke\kern -0.1em V}}}
\newcommand{\MeV}{\ensuremath{\mathrm{Me\kern -0.1em V}}}
\newcommand{\GeV}{\ensuremath{\mathrm{Ge\kern -0.1em V}}}
\newcommand{\TeV}{\ensuremath{\mathrm{Te\kern -0.1em V}}}
\newcommand{\ttbar}{\ensuremath{t\bar{t}}}
\newcommand{\mtop} {\ensuremath{M_{\mathrm{top}}}}
\newcommand{\GeVc}{\ensuremath{\mathrm{Ge\kern -0.1em V/c}}}
\newcommand{\GeVcc}{\ensuremath{\mathrm{Ge\kern -0.1em V/c}^2}}

\begin{document}

%\preprint{FERMILAB-PUB-12-486-E}

\title{Measurements of the Top-quark Mass and the $t\bar{t}$ Cross Section in the Hadronic $\tau$ + Jets Decay Channel at $\sqrt{s}=1.96 ~\TeV$}
%\author{}
%\thanks{}
%\affiliation{}


\affiliation{Institute of Physics, Academia Sinica, Taipei, Taiwan 11529, Republic of China}
\affiliation{Argonne National Laboratory, Argonne, Illinois 60439, USA}
\affiliation{University of Athens, 157 71 Athens, Greece}
\affiliation{Institut de Fisica d'Altes Energies, ICREA, Universitat Autonoma de Barcelona, E-08193, Bellaterra (Barcelona), Spain}
\affiliation{Baylor University, Waco, Texas 76798, USA}
\affiliation{Istituto Nazionale di Fisica Nucleare Bologna, $^{ee}$University of Bologna, I-40127 Bologna, Italy}
\affiliation{University of California, Davis, Davis, California 95616, USA}
\affiliation{University of California, Los Angeles, Los Angeles, California 90024, USA}
\affiliation{Instituto de Fisica de Cantabria, CSIC-University of Cantabria, 39005 Santander, Spain}
\affiliation{Carnegie Mellon University, Pittsburgh, Pennsylvania 15213, USA}
\affiliation{Enrico Fermi Institute, University of Chicago, Chicago, Illinois 60637, USA}
\affiliation{Comenius University, 842 48 Bratislava, Slovakia; Institute of Experimental Physics, 040 01 Kosice, Slovakia}
\affiliation{Joint Institute for Nuclear Research, RU-141980 Dubna, Russia}
\affiliation{Duke University, Durham, North Carolina 27708, USA}
\affiliation{Fermi National Accelerator Laboratory, Batavia, Illinois 60510, USA}
\affiliation{University of Florida, Gainesville, Florida 32611, USA}
\affiliation{Laboratori Nazionali di Frascati, Istituto Nazionale di Fisica Nucleare, I-00044 Frascati, Italy}
\affiliation{University of Geneva, CH-1211 Geneva 4, Switzerland}
\affiliation{Glasgow University, Glasgow G12 8QQ, United Kingdom}
\affiliation{Harvard University, Cambridge, Massachusetts 02138, USA}
\affiliation{Division of High Energy Physics, Department of Physics, University of Helsinki and Helsinki Institute of Physics, FIN-00014, Helsinki, Finland}
\affiliation{University of Illinois, Urbana, Illinois 61801, USA}
\affiliation{The Johns Hopkins University, Baltimore, Maryland 21218, USA}
\affiliation{Institut f\"{u}r Experimentelle Kernphysik, Karlsruhe Institute of Technology, D-76131 Karlsruhe, Germany}
\affiliation{Center for High Energy Physics: Kyungpook National University, Daegu 702-701, Korea; Seoul National University, Seoul 151-742, Korea; Sungkyunkwan University, Suwon 440-746, Korea; Korea Institute of Science and Technology Information, Daejeon 305-806, Korea; Chonnam National University, Gwangju 500-757, Korea; Chonbuk National University, Jeonju 561-756, Korea}
\affiliation{Ernest Orlando Lawrence Berkeley National Laboratory, Berkeley, California 94720, USA}
\affiliation{University of Liverpool, Liverpool L69 7ZE, United Kingdom}
\affiliation{University College London, London WC1E 6BT, United Kingdom}
\affiliation{Centro de Investigaciones Energeticas Medioambientales y Tecnologicas, E-28040 Madrid, Spain}
\affiliation{Massachusetts Institute of Technology, Cambridge, Massachusetts 02139, USA}
\affiliation{Institute of Particle Physics: McGill University, Montr\'{e}al, Qu\'{e}bec, Canada H3A~2T8; Simon Fraser University, Burnaby, British Columbia, Canada V5A~1S6; University of Toronto, Toronto, Ontario, Canada M5S~1A7; and TRIUMF, Vancouver, British Columbia, Canada V6T~2A3}
\affiliation{University of Michigan, Ann Arbor, Michigan 48109, USA}
\affiliation{Michigan State University, East Lansing, Michigan 48824, USA}
\affiliation{Institution for Theoretical and Experimental Physics, ITEP, Moscow 117259, Russia}
\affiliation{University of New Mexico, Albuquerque, New Mexico 87131, USA}
\affiliation{The Ohio State University, Columbus, Ohio 43210, USA}
\affiliation{Okayama University, Okayama 700-8530, Japan}
\affiliation{Osaka City University, Osaka 588, Japan}
\affiliation{University of Oxford, Oxford OX1 3RH, United Kingdom}
\affiliation{Istituto Nazionale di Fisica Nucleare, Sezione di Padova-Trento, $^{ff}$University of Padova, I-35131 Padova, Italy}
\affiliation{University of Pennsylvania, Philadelphia, Pennsylvania 19104, USA}
\affiliation{Istituto Nazionale di Fisica Nucleare Pisa, $^{gg}$University of Pisa, $^{hh}$University of Siena and $^{ii}$Scuola Normale Superiore, I-56127 Pisa, Italy}
\affiliation{University of Pittsburgh, Pittsburgh, Pennsylvania 15260, USA}
\affiliation{Purdue University, West Lafayette, Indiana 47907, USA}
\affiliation{University of Rochester, Rochester, New York 14627, USA}
\affiliation{The Rockefeller University, New York, New York 10065, USA}
\affiliation{Istituto Nazionale di Fisica Nucleare, Sezione di Roma 1, $^{jj}$Sapienza Universit\`{a} di Roma, I-00185 Roma, Italy}
\affiliation{Rutgers University, Piscataway, New Jersey 08855, USA}
\affiliation{Texas A\&M University, College Station, Texas 77843, USA}
\affiliation{Istituto Nazionale di Fisica Nucleare Trieste/Udine, I-34100 Trieste, $^{kk}$University of Udine, I-33100 Udine, Italy}
\affiliation{University of Tsukuba, Tsukuba, Ibaraki 305, Japan}
\affiliation{Tufts University, Medford, Massachusetts 02155, USA}
\affiliation{University of Virginia, Charlottesville, Virginia 22906, USA}
\affiliation{Waseda University, Tokyo 169, Japan}
\affiliation{Wayne State University, Detroit, Michigan 48201, USA}
\affiliation{University of Wisconsin, Madison, Wisconsin 53706, USA}
\affiliation{Yale University, New Haven, Connecticut 06520, USA}

\author{T.~Aaltonen}
\affiliation{Division of High Energy Physics, Department of Physics, University of Helsinki and Helsinki Institute of Physics, FIN-00014, Helsinki, Finland}
\author{B.~\'{A}lvarez~Gonz\'{a}lez$^z$}
\affiliation{Instituto de Fisica de Cantabria, CSIC-University of Cantabria, 39005 Santander, Spain}
\author{S.~Amerio}
\affiliation{Istituto Nazionale di Fisica Nucleare, Sezione di Padova-Trento, $^{ff}$University of Padova, I-35131 Padova, Italy}
\author{D.~Amidei}
\affiliation{University of Michigan, Ann Arbor, Michigan 48109, USA}
\author{A.~Anastassov$^x$}
\affiliation{Fermi National Accelerator Laboratory, Batavia, Illinois 60510, USA}
\author{A.~Annovi}
\affiliation{Laboratori Nazionali di Frascati, Istituto Nazionale di Fisica Nucleare, I-00044 Frascati, Italy}
\author{J.~Antos}
\affiliation{Comenius University, 842 48 Bratislava, Slovakia; Institute of Experimental Physics, 040 01 Kosice, Slovakia}
\author{G.~Apollinari}
\affiliation{Fermi National Accelerator Laboratory, Batavia, Illinois 60510, USA}
\author{J.A.~Appel}
\affiliation{Fermi National Accelerator Laboratory, Batavia, Illinois 60510, USA}
\author{T.~Arisawa}
\affiliation{Waseda University, Tokyo 169, Japan}
% ...

\collaboration{CDF Collaboration\footnote{With visitors from
$^a$Istituto Nazionale di Fisica Nucleare, Sezione di Cagliari, 09042 Monserrato (Cagliari), Italy,
$^b$University of CA Irvine, Irvine, CA 92697, USA,
$^c$University of CA Santa Barbara, Santa Barbara, CA 93106, USA,
$^d$University of CA Santa Cruz, Santa Cruz, CA 95064, USA,
$^e$Institute of Physics, Academy of Sciences of the Czech Republic, Czech Republic,
$^f$CERN, CH-1211 Geneva, Switzerland,
$^g$Cornell University, Ithaca, NY 14853, USA,
$^h$University of Cyprus, Nicosia CY-1678, Cyprus,
$^i$Office of Science, U.S. Department of Energy, Washington, DC 20585, USA,
$^j$University College Dublin, Dublin 4, Ireland,
$^k$ETH, 8092 Zurich, Switzerland,
$^l$University of Fukui, Fukui City, Fukui Prefecture, Japan 910-0017,
$^m$Universidad Iberoamericana, Mexico D.F., Mexico,
$^n$University of Iowa, Iowa City, IA 52242, USA,
$^o$Kinki University, Higashi-Osaka City, Japan 577-8502,
$^p$Kansas State University, Manhattan, KS 66506, USA,
$^q$Ewha Womans University, Seoul, 120-750, Korea,
$^r$University of Manchester, Manchester M13 9PL, United Kingdom,
$^s$Queen Mary, University of London, London, E1 4NS, United Kingdom,
$^t$University of Melbourne, Victoria 3010, Australia,
$^u$Muons, Inc., Batavia, IL 60510, USA,
$^v$Nagasaki Institute of Applied Science, Nagasaki, Japan,
$^w$National Research Nuclear University, Moscow, Russia,
$^x$Northwestern University, Evanston, IL 60208, USA,
$^y$University of Notre Dame, Notre Dame, IN 46556, USA,
$^z$Universidad de Oviedo, E-33007 Oviedo, Spain,
$^{aa}$CNRS-IN2P3, Paris, F-75205 France,
$^{bb}$Texas Tech University, Lubbock, TX 79609, USA,
$^{cc}$Universidad Tecnica Federico Santa Maria, 110v Valparaiso, Chile,
$^{dd}$Yarmouk University, Irbid 211-63, Jordan,
}}
\noaffiliation

\date{\today}

\begin{abstract}
We present the first direct measurement of the top-quark mass using 
$\ttbar$ events decaying in the hadronic $\tau$ + jets decay channel. 
Using data corresponding to an integrated luminosity of 2.2 $\fb$ 
collected by the CDF II detector in $p\bar{p}$ collisions at $\sqrt{s}=1.96 ~\TeV$ 
at the Fermilab Tevatron, we measure the $t\bar{t}$ cross section,
$\sigma_{t\bar{t}}$, and the top-quark mass, $\mtop$.  We extract $\mtop$ 
from a likelihood based on per-event probabilities calculated with 
leading-order signal and background matrix elements. We measure
$\sigma_{\ttbar} = 8.8 \pm 3.3 ~(\mathrm{stat}) \pm 2.2
~(\mathrm{syst}) ~\mathrm{pb}$ and $\mtop = 172.7 \pm 9.3
~(\mathrm{stat}) \pm 3.7 ~(\mathrm{syst}) ~\GeVcc$.
\end{abstract}

%\keywords{}

\maketitle

The mass of the top quark, $M_{\mathrm{top}}$, and the top-quark pair production
cross section, $\sigma_{t\bar{t}}$, have been extensively studied at
both the Fermilab Tevatron and the Large Hadron Collider at CERN 
\cite{ATLASref, CMSref, WAmass}. However, final states of the top-quark decay 
that include a tau lepton ($\tau$) are relatively unexplored due to the difficulty of 
identifying the tau and rejecting quantum chromodynamic (QCD) processes 
that can mimic its hadronic decay mode. The top quark and the tau 
belong to the heaviest third generation of standard model (SM) fermions and may 
play a special role in electroweak symmetry breaking. 
Discrepancies from the SM expectation in these unexplored decay channels 
could point to new top-quark sector physics.

In this Letter, we present the first direct measurement of 
the top-quark mass 
in the hadronic $\tau$ + jets decay channel ($\tau$ + jets) \cite{thesis}.   We 
measure $\sigma_{\ttbar}$ and $\mtop$ using data corresponding to 2.2 $\fb$ 
of integrated luminosity collected by the CDF II detector \cite{cdfdet} in 
$p\bar{p}$ collisions at $\sqrt{s}=1.96 ~\TeV$ at the Fermilab Tevatron. The 
D0 Collaboration previously measured $\sigma_{\ttbar}$ in the $\tau$ + jets 
decay channel with data corresponding to 1 $\fb$ of integrated luminosity to 
be $6.9 \pm 1.5$ pb \cite{d0tau}, assuming $\mtop = 170 ~\GeVcc$, and 
reached a signal purity of 52\%.

Assuming three generations in the SM and a unitary quark-mixing matrix, the top quark decays almost exclusively to a $W$ boson and $b$ 
quark. We select pair-produced top-quark events in which one of the $W$ bosons 
decays into a pair of light quarks and the other decays to a 
tau and a neutrino. This decay channel represents 15.2\% of the 
$t\bar{t}$ branching ratio and results in a final state with a tau, a neutrino, two 
$b$ quarks, and two light-flavor quarks ($u$, $d$, or $s$). Although the tau 
can decay leptonically to an electron ($e$) or muon ($\mu$) and a pair of 
neutrinos, these events are difficult to differentiate from electrons or muons from $W$ 
boson decays.  As a result, we select events with the tau decaying to a neutrino 
and a narrow jet of hadrons, which are usually charged and neutral pions, that correspond to 9.8\% of all $\ttbar$ decays. We use an artificial 
neural network (NN) to reduce the QCD multijet background contribution. 
The additional neutrino produced in the tau decay complicates the tau 
reconstruction.
To solve this, we adapt a missing mass 
calculator method \cite{tauColinearApprox} to the $\tau$ + jets topology to infer 
a unique solution for the neutrino four-momentum with sufficient precision to 
reasonably reconstruct the tau. We use a binned likelihood fit based on 
the predicted and observed number of events to measure $\sigma_{\ttbar}$. 
Then, to extract $\mtop$, we use a likelihood function built from signal and 
background probabilities calculated with the predicted differential cross 
sections for $\ttbar$ and $W$ + four-parton production, respectively.

The data used in this measurement are selected using a multijet online selection (trigger) that 
requires at least four calorimeter energy clusters with transverse energy
 \cite{bib:coord} ($E_T$) greater than $15 ~\GeV$ each and a total sum $E_T$ of all
clusters greater than $175 ~\GeV$ \cite{trigeff}. Jets are reconstructed by a 
cone algorithm that clusters energies in calorimeter towers within a fixed 
cone size of $\Delta R = 0.4$ \cite{bib:jetclu} where 
$\Delta R = \sqrt{\Delta \eta^2 + \Delta \phi^2}$ \cite{bib:coord}. 
In the offline analysis, events are required to have exactly four jets with 
$E_T> 20 ~\GeV$, missing transverse energy ($\met$) greater than $20 ~\GeV$, and 
a single hadronically-decaying tau selected as described below.  
Jet energies 
are corrected for nonlinearity of the detector response and multiple $p\bar{p}$ 
interactions within the bunch crossing \cite{bib:jetnim}.
One of 
the four jets must be identified as having originated from a $b$ quark 
($b$-tagged) using a secondary vertex finding algorithm \cite{secvtx}.  
Hadronically-decaying taus appear as narrow jets with an odd number 
of tracks and low neutral pion multiplicity. We select taus using a two-cones 
algorithm ~\cite{tauSel}. The inner cone defining the signal 
region has a size set to the lesser of $10^{\circ}$ (0.17 rad) and 
$(5 ~\GeV)/E_{\mathrm{cl}}$ rad, where $E_{\mathrm{cl}}$ is the energy of 
the calorimeter energy cluster associated with the tau candidate.  The 
second cone with a size of $30^{\circ}$ defines an isolation region outside 
of the signal cone. A tau must have one or three tracks in the signal region 
and no tracks in the isolation region. We require the $E_T$ of the tau energy cluster 
to exceed $20 ~\GeV$ and the $E_T$ of the visible tau to exceed $25 ~\GeV$, where visible 
refers to the combination of the track and neutral pion information.  
We require that the calorimeter energy in the isolation 
region be less than 10\% of the visible tau energy. Finally, we veto 
events with an identified electron or muon.  
  	
The dominant background for this analysis is high jet multiplicity QCD
events with a jet misidentified as a tau. To reduce this background, 
we develop a NN to distinguish between $t\bar{t} \rightarrow \tau$ + jets
and QCD multijet events.  
The NN is trained using 
QCD multijet events, obtained from data by selecting events with a tau 
candidate with at least one track in the isolation region and passing all 
other selection requirements, and $t\bar{t}$ events generated using the 
{\sc{pythia}} Monte Carlo generator \cite{pythia} coupled with a 
{\sc{geant}} \cite{bib:geant} based CDF II detector simulation \cite{bib:cdfsim}.
To properly account for tau polarization effects, the tau 
decays are modeled by the {\sc{tauola}} package \cite{tauola}.
The NN uses eight variables that exploit the topological differences 
between QCD multijet and $\ttbar$ events including $\met$ and 
the sum of the $E_T$ of various combinations of the tau and jets \cite{thesis}. 
After training the NN, we find good separation between QCD multijet and
$t\bar{t}$ events. Optimal signal significance, defined as the 
number of expected signal events divided by the square root of the total 
number of observed events,  is achieved by removing events associated to a NN 
value below 0.85. We initially select 162 events of which 41 events survive the 0.85 NN requirement.

\begin{figure}[htbp]
	\begin{center}
	\includegraphics[width=6cm]{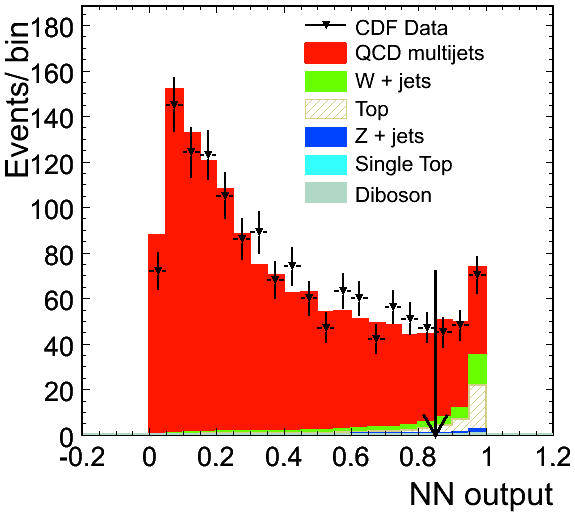}
		\caption{Fit to the NN output shape before applying the $b$-tagging requirement.  The arrow marks the lower bound on the signal region. \label{fig:fit_pretag}
		}
	\end{center}
	\end{figure}

Due to the difficulty in simulating QCD multijet events, $b$ quark 
tagging algorithms, and the production of heavy flavor quarks in association
with $W$ bosons, we estimate the background contributions with a data-driven 
approach similar to that described in Ref. \cite{secvtx}. We use the NN output 
distribution to fit the contributions of the signal and background processes to 
the data. This is done both before and after applying $b$-tagging requirements.  
Since most of the selected data events return a NN value below 0.7, 
the fit is dominated by events outside of the signal region.
We begin by calculating the contributions of the signal and each background 
process before the $b$-tagging requirement is applied. The $\ttbar$ and 
electroweak background contributions are determined from simulation.
Diboson, single top-quark, and $Z$ + jets production are modeled using 
{\sc{pythia}}, {\sc{madevent}} \cite{Madevent}, and {\sc{alpgen}} \cite{bib:Alpgen} 
respectively, with {\sc{pythia}} used for parton showering and underlying 
event generation. Each of these processes' contributions is set to its 
expectation based on its respective theoretical cross section 
\cite{BosonXsec,StopXsec}, the total integrated luminosity of the data, and the acceptance 
determined from simulation.  The $\ttbar$ contribution is modeled with 
{\sc{pythia}} and is similarly normalized using the next-to-next-to-leading order 
SM $\ttbar$ cross section prediction \cite{ttXsecTheory}. For all simulated 
events, a {\sc{geant}} based simulation is used to model the CDF II detector response.
With these contributions known, we determine the contributions from 
QCD multijet and $W$ + jets events by fitting the shape of the NN output
distribution for each component (with the previously calculated contributions 
fixed) to the data before applying the NN selection and $b$-tagging 
requirements.  The QCD multijet sample is selected from data as previously 
described while the $W$ + jets events are modeled with {\sc{alpgen}} similarly 
to $Z$ + jets events. We fit these distributions with a binned Poisson likelihood 
as seen in Fig. \ref{fig:fit_pretag}. The contribution of QCD multijet events in the 
signal region is calculated from this fit. All remaining events are assumed to come 
from $W$ + jets production.

For each process except QCD multijet events, the contribution after applying 
the $b$-tagging requirement is calculated by applying $b$-tagging efficiencies 
measured in Ref. \cite{secvtx} to the initially calculated contribution. Incorrect 
tagging of light quarks and tagging of the $b$ and $c$ quarks have
inherently different uncertainties. To properly estimate uncertainties associated 
with the contribution of $W$ + jets events, this contribution is divided into $W$ + 
light flavor ($W$+lf) and $W$ + heavy flavor ($Wb\bar{b}$, $Wc\bar{c}$, and 
$Wc$) parts with separately-estimated uncertainties. To calculate the contribution 
from QCD multijet events, we apply the $b$-tagging requirement to the QCD 
multijet sample. We then combine $\ttbar$ and the other background processes 
into a single sample with the relative contributions fixed to their calculated values. 
The NN output distributions of these two samples are then fit to the data selected 
with the $b$-tagging requirement, and the 
contribution of QCD multijet events in the signal region is derived from the result.
Each process's contribution, assuming $\sigma_{\ttbar} = 7.4$ pb and 
$\mtop = 172.5 ~\GeVcc$, is given in Table \ref{tab:tau_m2}.  Of the 41 selected 
events, we expect roughly 18 QCD multijet events and 18 $t\bar{t}$ events. 
From simulation studies, we estimate that 76.5 $\pm$ 0.5\% of the selected 
$\ttbar$ events correspond to a hadronic $\tau$ + jets final state. The other 
major contributions to the $\ttbar$ events are all-hadronic $\ttbar$ decays 
(12.3 $\pm$ 0.4\%) and $\ttbar \rightarrow e$ + jets (5.3 $\pm$ 0.3\%).

	\begin{table}
	\caption{Predicted number of selected $\tau$ + jets candidate events from each considered process after all selection requirements are applied assuming $\sigma_{t\bar{t}} = 7.4$ pb and $\mtop = 172.5 ~\GeVcc$. The uncertainty is a combination of statistical and systematic uncertainty. 
	\label{tab:tau_m2}}
	\begin{ruledtabular}
	\begin{tabular}{lc}
	Source & Number of events \\
	\hline
	Diboson           & $  0.19 \pm  0.01$  \\
	Single top quark  & $  0.16 \pm  0.01$  \\
	$Zb\bar{b}$       & $  0.29 \pm  0.04$  \\
	$Wb\bar{b}$       & $  0.6 \pm  0.5$  \\
	$Wc\bar{c}$       & $  0.3 \pm  0.3$  \\
	$Wc$              & $  0.2 \pm  0.1$  \\
	$W$+lf            & $  0.5 \pm  0.6$  \\
	QCD multijet     & $ 18.2 \pm  4.1$  \\
	Total bkgd        & $ 20.4 \pm 4.2 $ \\
	$\ttbar$          & $ 18.2 \pm  2.8$  \\
	Total predicted   & $ 38.6 \pm 5.0 $  \\
	Observed          & 41 \\
	\end{tabular}
		\end{ruledtabular}
	\end{table}
	
We measure $\sigma_{\ttbar}$ using a likelihood function based on a 
Poisson probability distribution comparing the number of observed ($N_o$) 
and predicted ($N_p$) events for a given $\sigma_{\ttbar}$ written as 
$L = e^{-N_p}N_p^{N_o}/(N_o!$). We consider the negative logarithm of this
function over values of $\sigma_{\ttbar}$ from 5 to 15 pb where 
$N_p$ is recalculated at each point with the fraction of QCD 
multijet events kept constant to the value calculated for 
$\sigma_{\ttbar} = 7.4$ pb.  The result is fit with a 2nd order
polynomial which is minimized to extract the central value and statistical
uncertainty.  The cross section value determined by the fit is 
$\sigma_{\ttbar} = 8.8 \pm 3.3 ~(\mathrm{stat})~\mathrm{pb}$.

The dominant sources of systematic uncertainty include the acceptance, selection
efficiencies, background estimate, and luminosity.  
For acceptance effects, we consider uncertainties on the 
jet energy scale (JES) \cite{bib:jetnim} (0.6 pb), 
parton showering models (0.5 pb), 
parton distribution functions (PDF) (0.5 pb),
initial and final state radiation (ISR, FSR) (0.5 pb), 
and color reconnection \cite{cr} (0.4 pb). 
We consider systematic uncertainties on 
the efficiency measurements from the $b$-tagging (0.4 pb), 
tau identification (0.2 pb), and trigger efficiency (0.1 pb) 
scale factors. The background systematics come from the $W$ + heavy flavor
scale factor uncertainty \cite{secvtx} (0.1 pb) and the QCD multijet
contribution, which is the dominant systematic uncertainty. 
We measure this uncertainty (1.8 pb) by comparing the NN 
output distribution shapes of the QCD multijet events and data 
dominated by QCD multijet events which are selected by removing 
the $\met$ requirement. 
Finally, the uncertainty on the integrated luminosity is 6\% \cite{lumi} (0.5 pb). 
Combining all these sources in quadrature results 
in the total systematic uncertainty of 2.2 pb, a 25\% uncertainty.  
We measure $\sigma_{\ttbar}$ assuming $\mtop = 172.5~\GeVcc$ 
to be $8.8 \pm 3.3 ~(\mathrm{stat}) \pm 2.2 ~(\mathrm{syst})
~\mathrm{pb}$, which is consistent with the next-to-next-to-leading order
SM prediction of $7.45^{+0.72}_{-0.63} ~\mathrm{pb}$ \cite{ttXsecTheory}.

We calculate $\mtop$ from a likelihood function based on probabilities 
corresponding to the signal and background hypothesis for each event. 
These probabilities are calculated from the differential cross section for 
$\ttbar$ and $W$ + four-parton production, respectively. The method 
uses a similar approach to the
previous measurement in the $e$ and $\mu$ +
jets decay channels \cite{meatv3}. The signal probability is based on
a $t\bar{t}$ leading-order matrix element which assumes $q\bar{q}$
production \cite{Mahlon:1995zn} and is calculated over 31 input mass
values ranging from 145 to 205 $\GeVcc$. Since it does not depend 
on $\mtop$, the background probability is 
calculated once for each event using a $W$ + four-parton matrix 
element from the {\sc{vecbos}} \cite{Vecbos:1991} generator. 

The tau decay adds an extra complication by introducing a second 
neutrino in the event.  We reconstruct this additional neutrino by adapting 
a method developed for the reconstruction of a resonance decaying 
to $\tau\tau$ \cite{tauColinearApprox} to the $\tau$ + jets topology. 
We find from simulation studies of $\ttbar \rightarrow \tau$ + jets events 
that the neutrino from the tau decay is nearly collinear with the hadronic 
components of the tau decay as their $\theta$ and $\phi$ angles 
tend to agree within 0.1 radians. Additionally, we find that the $\phi$ 
angle of the neutrino from the $W$ boson is within 1 radian of the $\phi$ 
direction associated with the $\met$. Simulation studies show that these 
statements hold true in greater than 99\% of events. We introduce a 
four-dimensional scan over the angles of both neutrinos restricting them 
to the above ranges. Assuming the neutrino mass is negligible and that 
the $W$ boson and tau have masses of $80.4 ~\GeVcc$ and 
$1.777 ~\GeVcc$, respectively, we completely solve for the four-vector 
of each neutrino for each set of angles in the scan. We then compare the 
predicted $x$- and $y$-components of the $\met$ from the neutrino solutions
to the measured $\met$ components with Gaussian probability functions 
and choose the set of angles that returns the greatest probability.  
This method accurately reconstructs the four-momentum of the neutrino 
from the tau decay, but it does not perform as well with the 
four-momentum of the neutrino from the $W$ boson decay.  Therefore, 
we use the result of this method only to determine the tau 
four-momentum in the event, while the neutrino from the $W$ boson decay is
reconstructed in the method as it would be for the $e$ or $\mu$ + jets 
channel \cite{meatv3} by assuming the $\ttbar$ system is produced with no $p_{T}$. 
Studies with simulated electron + jets events in which the electron energy is smeared 
to match the tau energy resolution show that any bias introduced by these methods 
is removed by the final calibration \cite{thesis}.

\begin{figure}[htbp]
	\begin{center}
	\includegraphics[width=6cm]{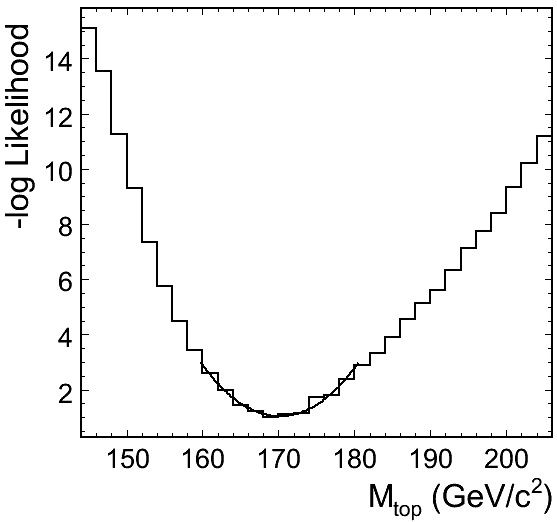}  
	\caption{Negative log of the top-quark mass likelihood as a function of $\mtop$ for all data events. The calibration functions have not yet been applied. \label{fig:massLike}}
	\end{center}
	\end{figure}

Each probability is calculated by integrating over the differential 
cross section for the appropriate process:
\begin{equation}
P = \frac{1}{\sigma}\int{d\sigma(\vec{y})f(\tilde{q_1})f(\tilde{q_2})W(\vec{x},\vec{y})d\tilde{q_1}d\tilde{q_2}},
\end{equation}
\noindent where $d\sigma$ is the differential cross section, $f$ is
the parton distribution function (PDF) for a quark with momentum
fraction of the incident proton $\tilde{q}$, $\vec{x}$ refers to observed 
quantities, $\vec{y}$ refers to parton level quantities, and 
$W(\vec{x},\vec{y})$ is the transfer function used to map $\vec{x}$ to 
$\vec{y}$ based on simulation studies.  
The event probability is a sum of the signal and 
background probabilities weighted by the signal and background 
fractions, respectively. To improve the statistical uncertainty on the 
$\mtop$ measurement, the likelihood function includes a Gaussian 
constraint on the background fraction set to $0.5 \pm 0.1$ from 
Table \ref{tab:tau_m2}. We evaluate the likelihood 
function for each of the 31 input top-quark masses and fit the result with 
a second-order polynomial to derive $\mtop$ and its statistical uncertainty. 
We calibrate the measurement on 21 simulated $\ttbar$ samples covering 
a mass range of 155 to 195 $\GeVcc$. The likelihood function and fit for 
the data before applying the calibration functions can be seen in 
Fig. \ref{fig:massLike}. We measure $\mtop$ to be $172.7
\pm 9.3 ~(\mathrm{stat}) ~\GeVcc$.

The largest systematic uncertainty comes from the JES and is
calculated to be 3.4 $\GeVcc$. We also consider systematic uncertainties
from the differences in parton showering models (0.5 \GeVcc), color 
reconnection (0.5 \GeVcc), ISR and FSR (0.3 \GeVcc), PDF's (0.1 \GeVcc), 
and the uncertainty on the fraction of $t\bar{t}$ pairs produced from 
$gg$ fusion (0.2 \GeVcc). The background fraction uncertainty is 
measured by shifting each background source within its uncertainty 
from Table \ref{tab:tau_m2} (0.5 \GeVcc). 
We consider uncertainties from different $b$-jet fragmentation models and
semileptonic branching ratios for jets from $b$ quarks as
well as shifts in the energy scale of these jets (0.4 \GeVcc).  We also
account for shifts from the tau energy scale (0.2 \GeVcc). The pileup
systematic uncertainty (1.0 \GeVcc) accounts for a known mismodeling in
the luminosity profile of the simulation. Uncertainty due to local non-linearity 
of the method and any assumptions used is estimated by 
shifting the calibration function within its uncertainty (0.2 \GeVcc). 
We take the remaining 0.14 $\GeVcc$ uncertainty on the fit of the mass residual 
(defined as the true mass subtracted from the measured mass) across all 
21 mass points as an uncertainty on the limited size of the simulation sample.
We find $\mtop$ to be 
$172.7 \pm 9.3 ~(\mathrm{stat})~ \pm 3.7 ~(\mathrm{syst})~\GeVcc$
 in agreement with the most recent Tevatron combination of 
 $173.18 \pm 0.94 ~\GeVcc$ \cite{WAmass}.

Using data corresponding to an integrated luminosity of 2.2 $\fb$, 
we have made the first direct measurement 
of the top-quark mass in $\ttbar$ events identified as decaying to 
a hadronic $\tau$ + jets topology. 
Assuming a top-quark mass of 172.5 $\GeVcc$, we find the $t\bar{t}$ pair production
cross section to be $8.8 \pm 3.3 ~(\mathrm{stat})\pm 2.2
~(\mathrm{syst+lumi})$ pb.  This value is consistent with the next-to-next-to-leading order SM
prediction \cite{ttXsecTheory} and recent measurements
\cite{xsec1}, including the D0 measurement in the same decay channel \cite{d0tau}. 
We measure the top-quark mass to be $172.7 \pm 9.3
~(\mathrm{stat}) \pm 3.7 ~(\mathrm{syst}) ~\GeVcc$ in agreement with
the Summer 2011 Tevatron top-quark mass combination of $173.18 \pm 0.94
~\GeVcc$ \cite{WAmass}.  
These measurements demonstrate 
that we can do complex analyses
with tau leptons even in a high jet multiplicity environment at hadron
colliders. This is particularly interesting at the LHC where new physics, 
e.g. SUSY, could preferentially lead to final states with tau leptons.

\begin{acknowledgments}
We thank the Fermilab staff and the technical staffs of the participating institutions for their vital contributions. This work was supported by the U.S. Department of Energy and National Science Foundation; the Italian Istituto Nazionale di Fisica Nucleare; the Ministry of Education, Culture, Sports, Science and Technology of Japan; the Natural Sciences and Engineering Research Council of Canada; the National Science Council of the Republic of China; the Swiss National Science Foundation; the A.P. Sloan Foundation; the Bundesministerium f\"ur Bildung und Forschung, Germany; the Korean World Class University Program, the National Research Foundation of Korea; the Science and Technology Facilities Council and the Royal Society, UK; the Russian Foundation for Basic Research; the Ministerio de Ciencia e Innovaci\'{o}n, and Programa Consolider-Ingenio 2010, Spain; the Slovak R\&D Agency; the Academy of Finland; and the Australian Research Council (ARC).
\end{acknowledgments}

\bibliography{topTau_prl_bib}

\begin{thebibliography}{28}
\bibitem [{bib()}]{bib:coord}%
  \BibitemOpen
  \bibinfo {note} {{C}DF uses a ($z$, $\phi$, $\theta$) coordinate system with
  the $z$-axis in the direction of the proton beam; $\phi$ and $\theta$ are the
  azimuthal and polar angle, respectively. The pseudorapidity is defined as
  $\eta = -\ln(\tan{\theta \over 2})$, and the transverse momentum and energy
  as $p_{\rm T} = p \sin \theta$ and $E_{\rm T} = E \sin \theta$, respectively.
  Missing transverse energy ($\met = |\metvec|$) is defined as $\metvec = -
  \sum_{i} E_{\rm T}^i {\bf{\hat{n}_i}}$ where ${\bf{\hat{n}_i}}$ is a unit
  vector in the transverse plane that points from the beam-line to the $i^{th}$
  calorimeter tower.}\BibitemShut {Stop}%
\end{thebibliography}

\end{document}